\documentclass{PoS}
\usepackage[latin1]{inputenc}
\usepackage{amsmath}
\usepackage{booktabs}

\title{Nucleon-pion-state contributions in the determination of the nucleon axial charge}

\ShortTitle{Nucleon-pion-state contributions to the nucleon axial charge}

\author{\speaker{Oliver B\"ar}\\
       Fakult\"at f\"ur Physik\\
       Universit\"at Bielefeld\\
       Universit\"atsstra{\ss}e 25\\
	D-33615 Bielefeld\\
	Germany \\
       E-mail: \email{obaer@physik.hu-berlin.de}}

\abstract{The nucleon-pion-state contributions to QCD 2- and 3-point functions used in the calculation of the nucleon axial charge are studied in chiral perturbation theory. For sufficiently small quark masses and large volumes the nucleon-pion states are expected to have smaller total energy than the single-particle excited states. To leading order in chiral perturbation theory the results do not depend on low-energy constants associated with the interpolating nucleon fields and apply to local as well as smeared interpolators. The nucleon-pion-state contribution is found to be at the few percent level.}

\FullConference{The 33rd International Symposium on Lattice Field Theory\\
		14 -18 July 2015\\
		Kobe International Conference Center, Kobe, Japan*}

\begin{document}

\section{Introduction}
In recent years quite a few collaborations started to pursue lattice QCD simulations with pion masses close to or at the physical value. For example, at this conference the PACS collaboration reported their latest results obtained on a $96^4$ lattice with $M_{\pi}\approx 147$MeV and $M_{\pi}L\approx 6$ \cite{UkitaLat2015}. Such simulations require no or only a short chiral extrapolation, so the uncertainties associated with this step are essentially eliminated. On the other hand, some complications get more severe the smaller the pion mass is. For instance, the signal-to-noise problem \cite{Lepage:1991ui} gets worse and prevents large euclidean time separations in many correlation functions. In addition, the smaller the pion mass the more pronounced is the contamination due to multi-particle-states in correlation functions. 

To discuss this second aspect consider the correlation functions needed to compute the nucleon axial charge $g_{\rm A}$. One usually computes the ratio $R(t,t') = C_3^A(t,t')/C_2(t)$ of the 3- and 2-pt functions
\begin{eqnarray}
 C^A_3(t,t')& = &\int d^3x\int d^3y \,\Gamma'_{k,\alpha\beta} \langle N_{\beta}(\vec{x},t) A^3_k(\vec{y},t') \overline{N}_{\alpha}(\vec{0},0)\rangle\,,\\
 C_2(t)& =& \int d^3x\,\Gamma_{\alpha\beta} \langle N_{\beta}(\vec{x},t)  \overline{N}_{\alpha}(\vec{0},0)\rangle\,.
\end{eqnarray}
$N_{\alpha},\overline{N}_{\beta}$ are nucleon interpolating fields placed at source time 0 and sink time $t$. $A^3_k$ denotes the third isospin component of the axial vector current at insertion time $t'$, and $\Gamma', \Gamma$ are appropriately chosen Dirac matrices (see Ref.\ \cite{Bhattacharya:2015wna}, for example). Using the spectral decomposition one finds for $t\gg t'\gg 0$ 
\begin{equation}
R(t,t') \approx g_A +b_1 e^{-\Delta E_1 (t-t')} + \tilde{b}_1 e^{-\Delta E_1 t'} + c_1 e^{-\Delta E_1 t}+ \ldots\,.
\end{equation}
The deviations from a constant are due to excited states with the same quantum numbers as the nucleon. The exponential suppression is governed by the energy gap $\Delta E_1 = E_1 - M_N$ where $E_1$ is the energy of the first excited state. The nature of this state depends strongly on the pion mass and the extent $L$ of the finite spatial box realized in the simulation. For pion masses close to the physical value and for sufficiently large $L$ one expects the nucleon-pion state with back-to-back momentum $|\vec{p}|=2\pi/L$ to be the one with the smallest energy, $E_1\approx E_N + E_{\pi}$. Even the three particle state with a nucleon and two pions at rest has a smaller energy than the first single-particle excited state. Assuming the aforementioned values $M_{\pi}\approx 147$MeV and $M_{\pi}L\approx 6$ one expects five nucleon-pion states with an energy between $M_N$ and $1.5M_N$.		

The size of the excited state contributions does depend on the size of the coefficients $b_1,\tilde{b}_1,c_1$ too. These coefficients are ratios of matrix elements involving the excited states. As pointed out in Ref.\ \cite{Bar:2012ce} these coefficients can be computed in chiral perturbation theory (ChPT). This idea is not new and has been independently put forward in \cite{Tiburzi:2009zp,Tiburzi:2015tta,Tiburzi:2015sra} using heavy baryon ChPT. Here we report our results using the covariant ChPT formulation, taking into account the chiral expansion of the nucleon interpolating fields. Moreover, we compare the size of the nucleon-pion-state contributions  in three different methods to compute the axial charge.  

\section{Setup: Covariant Baryon ChPT}
The framework for our calculation is the covariant formulation of Baryon ChPT \cite{Gasser:1987rb,Becher:1999he} (see also the monograph \cite{Scherer:2012xha}). Both the chiral effective Lagrangian and the expressions for the vector and axial vector currents can be taken from the literature. The chiral expressions for the nucleon interpolating fields are less known but were derived in \cite{Wein:2011ix,Bar:2015zwa}. Starting point are the familiar local nucleon interpolating fields in QCD, $N_1=(\tilde{q}q)q$ and $N_2=(\tilde{q}\gamma_5q)\gamma_5 q$, with $q$ denoting the quark doublet containing the up and down quark, and $\tilde{q} = q^{\rm T} C\gamma_5(i\sigma_2)$. Based on the transformation properties under chiral symmetry and parity \cite{Nagata:2008zzc} these interpolating fields are mapped to ChPT by writing down the most general expression in Baryon ChPT that has the same transformation properties. The resulting expression is then expanded in powers of the pion fields as usual. 

Local interpolating fields are rarely used in lattice simulations. Instead, smeared interpolating fields made from smeared quark fields $q_{\rm sm}$ are usually employed, where  $q_{\rm sm}(x)= \int d^4y K(x-y) q(y)$ with some gauge covariant kernel $K$. The details of the kernel depend on the particular smearing (e.g.\ Gaussian smearing \cite{Gusken:1989ad,Alexandrou:1990dq}, the gradient flow \cite{Luscher:2013cpa}, etc.) For what matters here only two properties are important: i) the kernel is diagonal in spinor space and ii) it is essentially zero for distances $|x-y|$ larger than $R_{\rm smear}$, a ``smearing radius''. In that case the smeared interpolating nucleon fields have the same transformation properties as their unsmeared counterparts. If, in addition, the smearing radius is much smaller than the Compton wave length of the pion the smeared interpolating nucleon fields 
are mapped onto the same pointlike expressions in ChPT as the local ones, the only difference is the different low-energy constants (LECs) in these expressions.     

\begin{figure}[t]
\begin{center}
 \includegraphics[scale=0.4]{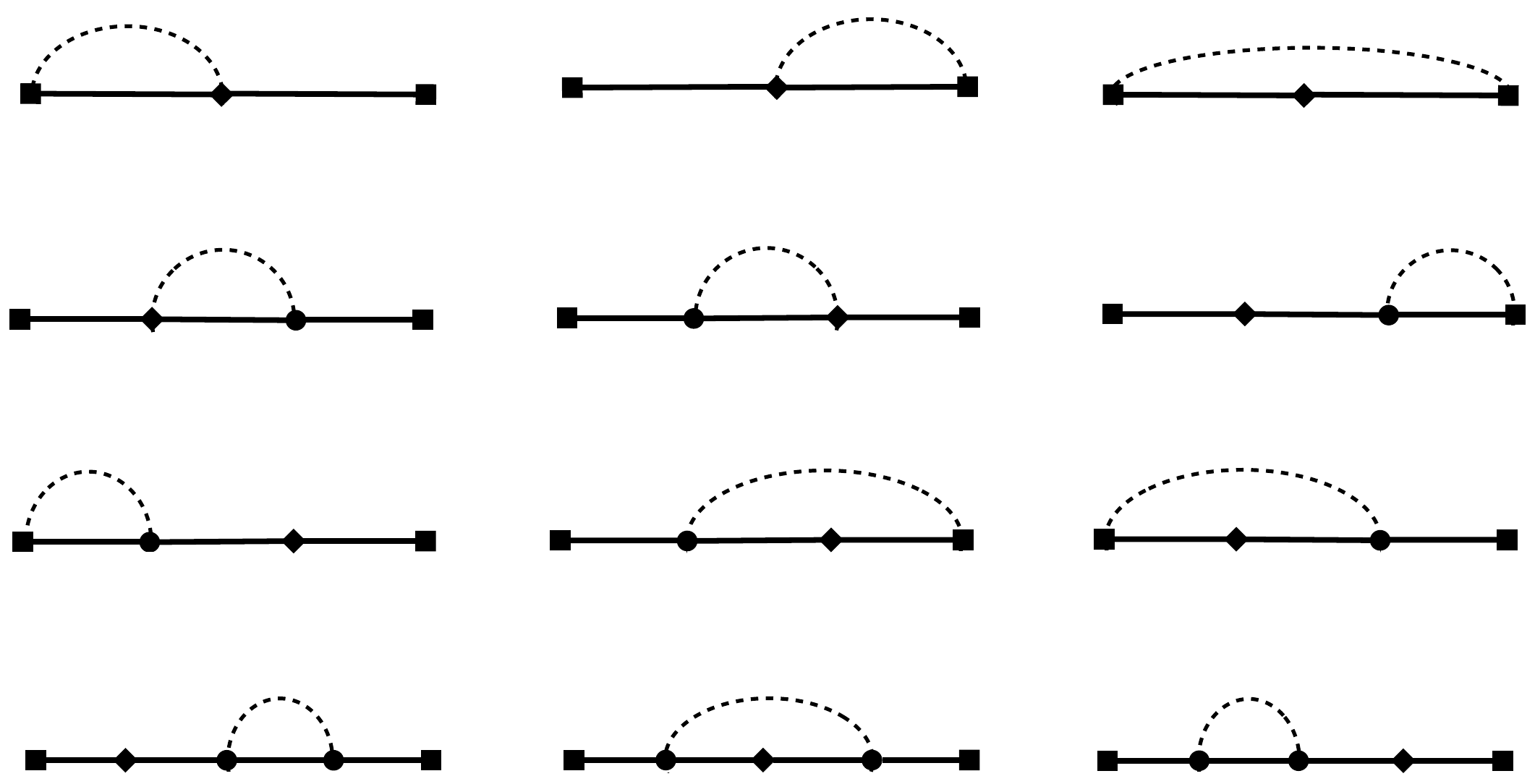}
\caption{Feynman diagrams for the 3-pt function. Solid and dotted lines correspond to nucleon and pion propagators, respectively. Squares, diamonds and circles stem from the interpolating nucleon fields, the current and the interaction vertex.}
\label{diagrams}
\end{center}
\end{figure}

The calculation of the correlation functions in ChPT is straightforward. To LO it involves sixteen diagrams for the 3-pt function, depicted in fig.\ \ref{diagrams}. Four diagrams contribute to the 2-pt function, which was already computed in \cite{Bar:2015zwa}. A finite box with spatial extent $L$ and periodic boundary conditions was assumed. The time extent was taken to be infinite, for simplicity. In addition, we assumed  isospin symmetry.

\section{Nucleon-pion contribution to $g_{\rm A}$}
Taking into account the nucleon-pion state contribution the result for the ratio $R(t,t')$ assumes the general form
\begin{equation}\label{RatioR}
R(t,t') \approx g_A +\sum_{p_n}\left( b_n e^{-\Delta E_n (t-t')} + \tilde{b}_n e^{-\Delta E_n t'}+c_n e^{-\Delta E_n t}\right).
\end{equation}
The sum runs over the discrete momenta allowed by the periodic boundary conditions. The energy gaps are given by $\Delta E_n = E_N(p_n)+E_{\pi}(p_n) - M_N$.  To leading order (LO), the order we are working here, we find $b_n=\tilde{b}_n$. This is expected if the same interpolating field is used at source and sink. Deviations from $b_n=\tilde{b}_n$ show up at higher order in the chiral expansion.

The coefficients $b_n,\tilde{b}_n,c_n$ are dimensionless and can be written as functions of the following dimensionless parameters: $fL, \,g_{\rm A},\, M_{\pi}/M_N$ and $M_{\pi}L$. Two LO LECs enter, $f$ and $g_{\rm A}$, but the coefficients do not depend on the LECs associated with the nucleon interpolating fields. The reason is chiral symmetry. It dictates that these LECs are overall factors in both the 3-pt and 2-pt function, thus they cancel in the ratio $R$. Therefore, LO ChPT makes a rather definite prediction for the ratio. For our purposes we can approximate $f$ and $g_{\rm A}$ by the experimentally well-known measured values for the pion decay constant and the axial charge. Once this is done the coefficients depend on two unknowns only, $M_{\pi}/M_N$ and $M_{\pi}L$.

\begin{figure}[tbp]
\begin{center}
 \includegraphics[scale=0.43]{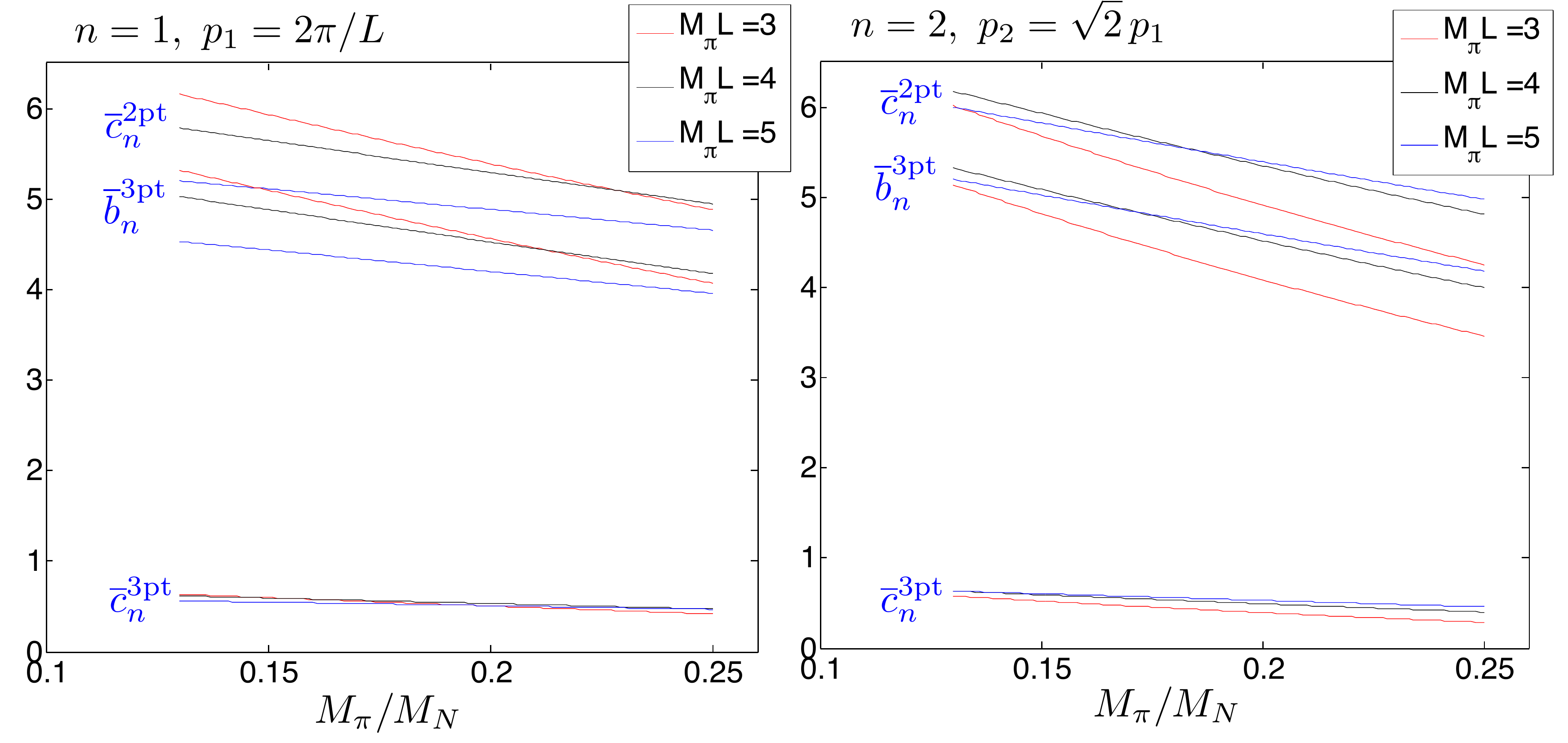}
\caption{The coefficients $\overline{b}_n^{\rm 3pt}, \overline{c}_n^{\rm 3pt}$ and $\overline{c}_n^{\rm 2pt}$ as a function of $M_{\pi}/M_N$ and $M_{\pi}L$ for the lowest two momentum states ($n=1$ and $n=2$).}
\label{figcoeff}
\end{center}
\end{figure}

It is useful to separate a trivial prefactor and the non-trivial ChPT result in the coefficients by writing
\begin{equation}\label{barredcoeff}
b_n = \frac{m_n}{16(fL)^2 E_{\pi}L}\overline{b}_n^{\rm 3pt}\,,\quad c_n = \frac{m_n}{16(fL)^2 E_{\pi}L}(\overline{c}_n^{\rm 3pt} - \overline{c}_n^{\rm 2pt})\,.
\end{equation}
$m_n$ here is the multiplicity of momentum states counting the number of states with the same energy ($m_1=6,m_2=12$, see Ref.\ \cite{Colangelo:2003hf}). Note that the coefficient $c_n$ has two contributions. One ($\overline{c}_n^{\rm 3pt}$) stems directly from the 3-pt function, the second one ($\overline{c}_n^{\rm 2pt}$) has its origin in the denominator of the ratio $R$, i.e.\ the 2-pt function. 

In figure \ref{figcoeff} the coefficients for the two smallest spatial momenta are plotted as functions of $M_{\pi}/M_N$ for three values of $M_{\pi}L$. The dependence on these two variables is very mild. Qualitatively the same behaviour is found for the coefficients with $n>2$ too. The following observations can be made: Firstly, all coefficients are positive. Secondly, $\overline{c}_n^{\rm 3pt}$ is roughly a factor ten smaller than $\overline{c}_n^{\rm 2pt}$, hence $\overline{c}_n^{\rm 3pt}-\overline{c}_n^{\rm 2pt}\approx - \overline{c}_n^{\rm 2pt}<0$. And finally, $\overline{b}_n^{\rm 3pt}\approx \overline{c}_n^{\rm 2pt}$. All this implies that the coefficients in the ratio \ref{RatioR} are of the same order of magnitude but have opposite sign, $b_n=\tilde{b}_n \approx - c_n$. As a consequence, the two nucleon-pion-state contributions to the ratio compensate each other to a large extent.

\begin{figure}[p]
\begin{center}
 \includegraphics[scale=0.45]{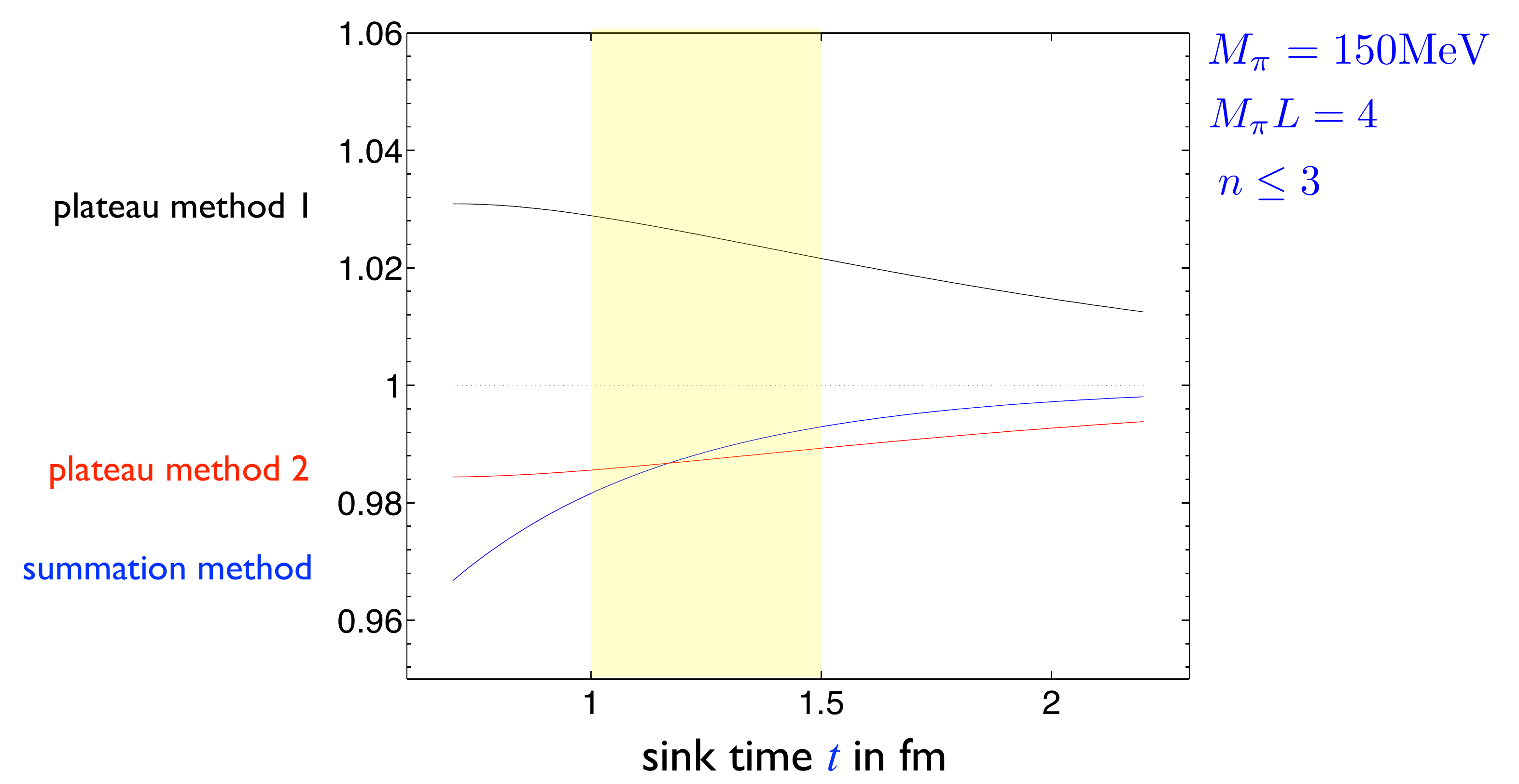}
\caption{Ratio of the estimate and the asymptotic value $g_{\rm A}$ as a function for three different methods to obtain the axial charge (see main text). }
\label{comparison1}
\end{center}
\end{figure}

In the following we compare three different ways to obtain an estimate for $g_{\rm A}$. The first one takes $g_{\rm A} \approx R(t,t'=t/2)$ since the excited state contribution is minimal if the insertion time is half the sink time. This ``midpoint method'' is essentially equivalent to the usual plateau method, and we refer to it by plateau method 1. The summation method \cite{Maiani:1987by} sums $R(t,t')$ over all $t'$ between source and sink and extracts the axial charge from the term linear in $t$. This approximation for $g_{\rm A}$ is equivalent to $R$ but with the $b_n$-, $\tilde{b}_n$-contribution dropped. Finally, one can obtain the axial charge also by replacing the 2-pt function in $R$ with the 3-pt function of the vector current (both 3-pt functions with $t'=t/2$), and we refer to this as plateau method 2. 

\begin{figure}[p]
\begin{center}
 \includegraphics[scale=0.4]{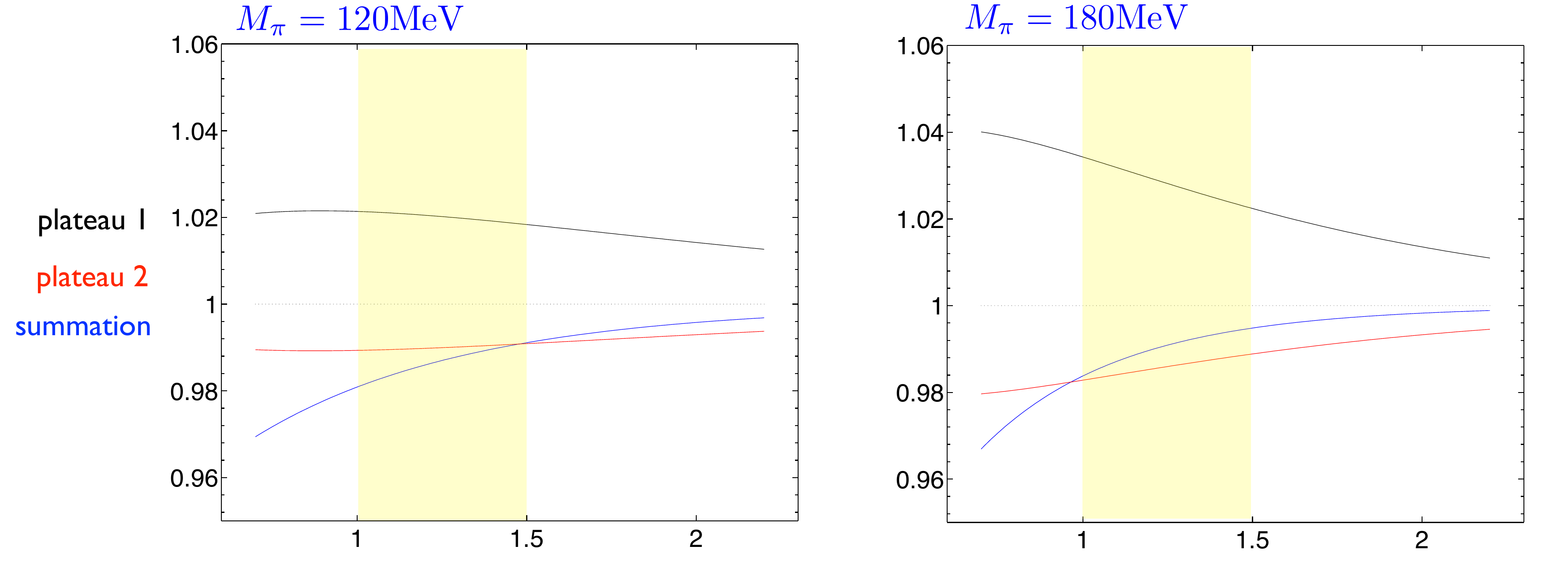}\newline
  \includegraphics[scale=0.4]{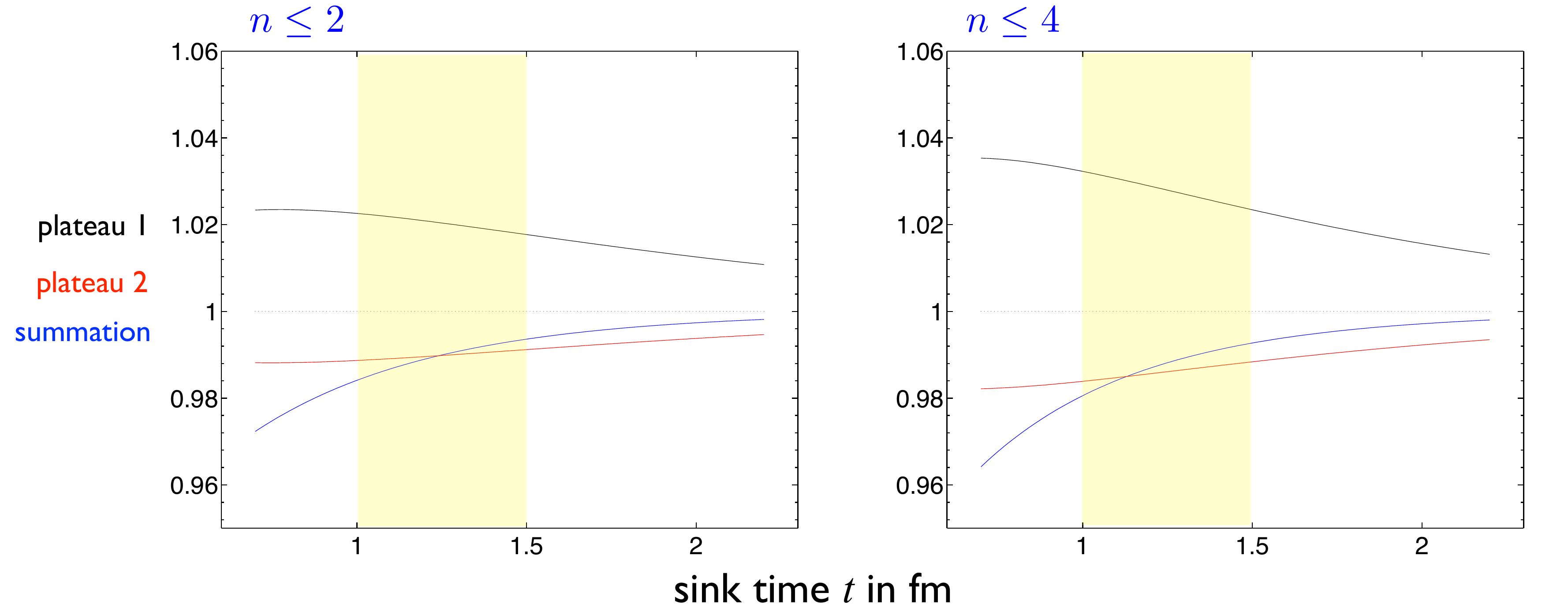}
\caption{Same as in figure 3 but with different pion masses (upper two panels) and with different number of nucleon-pion-state contributions included (lower two panels). }
\label{comparison2}
\end{center}
\end{figure}

All three methods give estimates for the axial charge that depend on the sink time $t$. Figure \ref{comparison1} shows this 
 dependence for a pion mass $M_{\pi}=150$MeV and $M_{\pi}L=4$. Plotted is the estimate divided by $g_{\rm A}$, so without the excited state contaminations all three methods give 1.  The three curves show the result including the nucleon-pion contribution taking into account the first three momentum states, i.e.\ $n\le3$. The yellow region $1{\rm fm} \le t \le 1.5 {\rm fm}$ highlights the range of sink time that seems feasible in numerical simulations.
 The overall conclusion we can draw from this figure is that the nucleon-pion state contribution is at the few percent level. For the standard plateau method we find a 2-3\% shift upwards, while the other two methods show a slightly smaller shift downwards. The results are fairly stable under changes of the pion mass  and the number of states included in the contamination, see fig.\ \ref{comparison2}.

Not obvious from these figures is the partial compensation of the nucleon-pion-state contribution mentioned before. For example, for a sink time $t=1{\rm fm}$ in fig.\ \ref{comparison1} we read off a +3\% correction for plateau method 1. These 3\% are essentially the difference (5-2)\%, where the 5\% and 2\% stem from the $b_n$- and $c_n$-contribution in $R$, respectively. If the coefficient $\overline{c}_n^{\rm 3pt}$ were approximately equal to  $\overline{c}_n^{\rm 2pt}$ (and not a factor ten smaller) the nucleon-pion-state contribution would be 5\%, so almost a factor 2 larger.

Another consequence of the partial compensation is a fairly flat sink time dependence for small pion masses. In the upper left panel of figure \ref{comparison2}, for example, the two plateau methods show an almost constant nucleon-pion-state contribution over the yellow sink time region. In contrast, the summation method still shows an exponential decay, because the nucleon-pion contribution for this method consists of the $c_n$-contribution only and no partial compensation takes place.  Whether this difference can be resolved in practice is questionable, given that very precise data would be needed. Still, the results demonstrate that an increase in sink time not necessarily leads to a smaller excited state contribution, at least not for the sink times accessible in practice.

The results presented here were obtained in a LO calculation. This obviously raises the questions about higher oder corrections.
These are in principle straightforward to compute. However, at higher order additional LECs are expected to enter that may be poorly known and the higher order corrections are probably less predictive than the LO results discussed here. 
Another source of uncertainty is the mapping of smeared interpolating fields to pointlike fields in ChPT. Naively this may result in an error of ${\rm O}(M_{\pi}R_{\rm smear})$, which is about 30\% for a smearing radius of 0.3fm and a pion mass less than 200MeV. Therefore, it would not be surprising if the LO result given here has an error of about 100\%. For a more precise estimate the higher order corrections need to be calculated. Still, the sign and the order of magnitude for the nucleon-pion state contribution seem to be rather solid results. 

As a final remark we mention that the multi-particle-state contamination with more the one pion can be computed in an analogous way. However, the impact of these states is even smaller than the nucleon-pion-states considered here. For example, the 3-particle $N\pi\pi$-state contribution is suppressed by an additional factor $[2(fL)^2E_{\pi}L]^{-1}$ and too small to play a role in practice.

\section{Summary and outlook}
ChPT provides estimates for the nucleon-pion-state contribution in nucleon correlation functions. In case of the nucleon axial charge this contribution is at the few percent level at LO in the chiral expansion. Analogous calculations for other observables determined by nucleon 3-point functions are straightforward. Particularly interesting are the tensor and scalar nucleon charges, the electromagnetic form factors and the quark momentum fraction in the nucleon.

\end{document}